\documentclass[12pt]{iopart}
\usepackage{graphicx}
\usepackage{citesort}
\usepackage{epstopdf}

\begin{document}

\title{The meeting problem in the quantum random walk}

\author{ M. \v Stefa\v n\'ak$^{(1)}$, T. Kiss$^{(2)}$, I.
Jex$^{(1)}$ and B. Mohring$^{(3)}$}

\address{$^{(1)}$~Department of Physics, FJFI \v CVUT, B\v
rehov\'a 7, 115 19 Praha 1 - Star\'e M\v{e}sto, Czech Republic}

\address{$^{(2)}$~Department of Nonlinear and Quantum Optics, Research
Institute for Solid State Physics and Optics, Hungarian Academy of
Sciences, Konkoly-Thege u.29-33, H-1121 Budapest, Hungary}

\address{$^{(3)}$~Department of Quantum Physics, University of Ulm, D-89069
Ulm, Germany}

\date{today}

\begin{abstract}
We study the motion of two non-interacting quantum particles
performing a random walk on a line and analyze the probability that
the two particles are detected at a particular position after a
certain number of steps (meeting problem). The results are compared
to the corresponding classical problem and differences are pointed
out. Analytic formulas for the meeting probability and its asymptotic
behavior are derived. The decay of the meeting probability for
distinguishable particles is faster then in the classical case, but not 
quadratically faster. Entangled initial states and the bosonic or 
fermionic nature of the walkers are considered.
\end{abstract}


\submitto{\JPA}

\maketitle

\section{Introduction}

Random walks are a long studied problem not only in classical
physics but also in many other branches of science \cite{crw}.
Random walks help to understand complex systems, their dynamics and
the connection between dynamics and the underlying topological
structure. More recently, quantum analogues \cite{Aharonov,Kempe} of
the classical random walks attracted considerable attention.

Quantum walks have been introduced in the early nineties by
Aharonov, Davidovich and Zagury \cite{Aharonov}. Since that the
topic attracted considerable interest. The continuing attraction of
the topic can be traced back to at least two reasons. First of all,
the quantum walk is, as a topic, of sufficient interest on its own
right because there are fundamental differences compared to the
classical random walk. Next, quantum walks offer quite a number of
possible applications. One of the best known is the link between
quantum walks and quantum search algorithms which are superior to
their classical counterparts \cite{Shenvi,Childs}. Another goal is
to find new, efficient quantum algorithms by studying various types
of quantum walks. Let us point out that from a quantum mechanical
point of view there is no randomness involved in the time evolution
of a quantum walk. The evolution of the walker is determined
completely by a unitary time evolution. This applies to the two most
common forms: the discrete \cite{Aharonov} and the continuous time
\cite{Childs} quantum walks. For a review on quantum walks see for
instance \cite{Kempe}.

Several aspects of quantum walks have been analyzed. First of all
quite much attention was paid to the explanation of the asymptotics
of the unusual walker probability distribution using various
approaches \cite{qwol,knight,carteret1,carteret}. Next, attention
was paid to the explanation of the unusual probability distribution
as a wave phenomenon \cite{knight,Jeong}. This question is of
particular interest as there are several proposals how to realize
quantum walks \cite{NMR,Agarwal,Hillery,Jeong,Galton,Trav,Dur,ion},
in particular proposals using optical elements as the basic blocks
for the quantum walks \cite{Hillery,Jeong}. Among the first ones was
the optical implementation of a Galton board \cite{Galton}. An ion
trap proposal as well as an neutral atom implementation was put
forward few years later \cite{Trav,Dur,ion}. The neutral atom
proposal lead to a real experiment in the year 2003 \cite{Man}.

Apart from the studies of elementary random walks its
generalizations have been studied \cite{ctw}. Generalizations of
random walks to higher dimensions \cite{2dqw1} have been put forward
and differences to the simpler models pointed out. Next the effect
of randomness in optical implementations on quantum walks was
analyzed and its link to localization pointed out \cite{loc}. Among
further generalizations also the behavior of more than one walker
(particle) in networks realizing random walks has been studied
\cite{lindberg,omar}. The aim of the present paper is to add to this
line of studies. We wish to study the evolution of two walkers
performing a random walk. The evolution of each of the two walkers
is subjected to the same rules. One of the interesting questions
when two walkers are involved is to clarify how the probability of
the walkers to meet changes with time (or number of steps taken in
walk). Because the single quantum walker behavior differs from its
classical counterpart it has to be expected that the same will apply
to the situation when two walkers are involved. Interference,
responsible for the unusual behavior of the single walker should
play also a considerable role when two walkers will be involved. The
possibility to change the input states (in particular the
possibility to choose entangled initial walker states) adds another
interesting point to the analysis. In the following, we will study
the evolution of the meeting probability for two walkers. We point
out the differences to the classical case and discuss the influence
of the input state on this probability.

The paper is organized as follows - first we make a brief review of
the concept of the discrete time quantum walk on an infinite line
and its properties. Based on this we generalize the two particle
random walk for both distinguishable and indistinguishable walkers
and define the meeting problem. This problem is analyzed in section
3. The asymptotic behavior of the meeting probability is derived.
Further the effect of entanglement and indistinguishability of the
walkers is examined. In the conclusions we summarize the obtained
results and discuss possible development. Finally, in the appendix
the properties of the meeting problem in the classical random walk
are derived.

\section{Description of the walk}

We first briefly summarize the description of the quantum random
walk, for more details see e.g. \cite{Kempe}. We consider a coined
random walk on an infinite line. The Hilbert space of the particle
$\mathcal{H}$ consists of the position space $\mathcal{H}_P$ with
the basis $\left\{|m\rangle|m\in\mathrm{Z}\right\}$, which is
augmented by a coin space
\begin{equation}
\mathcal{H}_C = \left\{|L\rangle,|R\rangle\right\}.
\end{equation}
The particle moves on the grid in discrete time steps in dependence
on the coin state, the operator which induces a single displacement
has the form
\begin{equation}
S = \sum_{m}|m-1\rangle\langle m|\otimes|L\rangle\langle
L|+\sum_{m}|m+1\rangle\langle m|\otimes|R\rangle\langle R|.
\end{equation}
A single step of the particle consists of a rotation of the coin
state given by an arbitrary unitary matrix $C$ and the conditional
displacement $S$. The time evolution operator $U$ describing one
step of the random walk takes the form
\begin{equation}\label{U1}
U = S(I\otimes C).
\end{equation}
If the initial state of the walker is $|\psi(0)\rangle$, then after
$t$ steps its state will be given by the successive application of
$U$ on the initial state
\begin{equation}\label{te}
|\psi(t)\rangle = U^t|\psi(0)\rangle.
\end{equation}
The probability distribution generated by such a random walk is
given by
\begin{equation}
P(m,t) = |\langle m,L|\psi(t)\rangle|^2+|\langle
m,R|\psi(t)\rangle|^2.
\end{equation}
In our paper we concentrate on a particular case when the rotation
of the coin is given by the Hadamard transformation
\begin{equation}
H = \frac{1}{\sqrt{2}}\left(%
\begin{array}{cc}
  1 & 1 \\
  1 & -1 \\
\end{array}%
\right),
\end{equation}
since this is the most studied example of an unbiased random walk.
For quantum walks with arbitrary unitary coin see e.g.
\cite{Tregenna}.

We will describe the wave-function of the walker at time $t$ by the
set of two-component vectors of the form
\begin{equation}
\psi(m,t)=\left(%
\begin{array}{c}
  \psi_L(m,t) \\
  \psi_R(m,t) \\
\end{array}%
\right),
\end{equation}
where $\psi_L(m,t)$ ($\psi_R(m,t)$) is the probability amplitude
that the walker is at time $t$ on the site $m$ with the coin state
$|L\rangle$ ($|R\rangle$). The wave-function thus has the form
\begin{equation}
|\psi(t)\rangle = \sum_m\left(
\psi_L(m,t)|m,L\rangle+\psi_R(m,t)|m,R\rangle\right).
\end{equation}

Throughout the text we will use symbols $|\psi^{L(R)}(t)\rangle$,
$\psi^{L(R)}_{i,j}(m,t)$ for the vectors and the probability
amplitudes, under the assumption that the initial state of the
walker was
\begin{equation}
|\psi^{L(R)}(0)\rangle=|0,L(R)\rangle,
\end{equation}
similarly $P^{L(R)}(m,t)$ will be the corresponding probabilities.

From the time evolution of the wave-function (\ref{te}) follows the
dynamics of the two-component vectors $\psi (m,t)$
\begin{equation}\label{te2}
\psi(m,t+1) = \frac{1}{\sqrt{2}}\left(%
\begin{array}{cc}
  0 & 0 \\
  1 & -1 \\
\end{array}%
\right)\psi(m-1,t)+\frac{1}{\sqrt{2}}\left(%
\begin{array}{cc}
  1 & 1 \\
  0 & 0 \\
\end{array}%
\right)\psi(m+1,t).
\end{equation}
Thus the description of the time evolution of the walker reduces to
a set of difference equations. Nayak and Vishwanath in \cite{qwol}
have found the analytical solution of (\ref{te2}) and derived the
asymptotic form of the probability distribution. Before we proceed
with the generalization of the quantum walk for two particles we
will summarize the properties of the single walker probability
distribution derived in \cite{qwol}, which we will use later for the
estimation of the meeting probability. According to \cite{qwol} the
probability distribution of one walker is almost uniform in the
interval $(s-t/\sqrt{2},s+t/\sqrt{2})$, where $s$ is the initial
position of the walker, with the value $\approx 1/t$. Around the
points $s\pm t/\sqrt{2}$ are peaks of width of the order of
$t^{1/3}$ and the probability is approximately $t^{-2/3}$. Outside
this region the probability decays faster then any inverse
polynomial in $t$ and therefore we will neglect this part. We will
also use the slow oscillating part of the walker probability
distribution derived in \cite{qwol} which has the form
\begin{equation}\label{pslow}
P^{L(R)}_{slow}(x,t) = \frac{2}{\pi
t\left(1\pm\frac{x}{t}\right)\sqrt{1-\frac{2x^2}{t^2}}},
\end{equation}
for the case that the initial coin state was $L(R)$.

The main difference between classical and quantum random walk on a
line is the shape of the probability distribution. Due to the
interference effect the quantum walker spreads quadratically faster
than the classical one and its variance goes like $\sigma\sim t$, in
contrast with the result for the classical case
$\sigma\sim\sqrt{t}$.

To visualize the properties of the quantum walk we plot in figure 1
the probability distribution and the slow-varying estimation. The
initial conditions of the coin were chosen to be
$\frac{1}{\sqrt{2}}(|L\rangle+i|R\rangle)$, which leads to an
unbiased distribution. The slow-varying estimation for the symmetric
probability distribution is given by
\begin{equation}\label{pslowsym}
P^{sym}_{slow}(x,t) =
\frac{1}{2}\left(P^L_{slow}(x,t)+P^R_{slow}(x,t)\right) =
\frac{2}{\pi
t\left(1-\frac{x^2}{t^2}\right)\sqrt{1-\frac{2x^2}{t^2}}}.
\end{equation}
\begin{figure}
\begin{center}
\includegraphics[width=13cm]{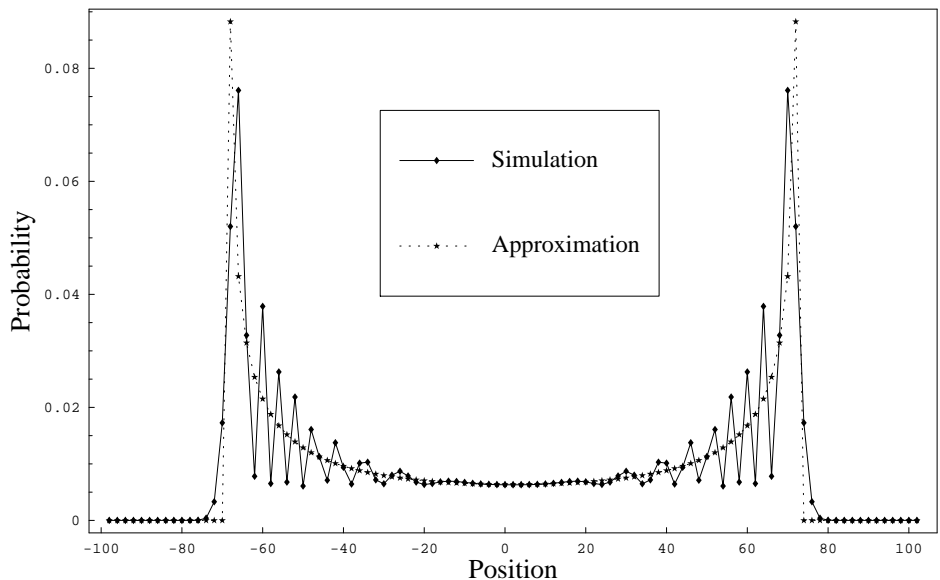}
\end{center}
\caption{Probability distribution of a quantum random walker after
100 steps and the slow-varying estimation.}
\end{figure}
Using this description of the random walk we now proceed to the
analysis of the random walk for two non-interacting distinguishable
and indistinguishable particles.

\subsection{Distinguishable walkers}

The Hilbert space of the two walkers is given by a tensor product of
the single walker spaces, i.e.
\begin{equation}
\mathcal{H} =
(\mathcal{H}_P\otimes\mathcal{H}_C)_1\otimes(\mathcal{H}_P\otimes\mathcal{H}_C)_2.
\end{equation}
Each walker has its own coin which determines his movement on the
line. Since we assume that there is no interaction between the two
walkers they evolve independently and the time evolution of the
whole system is given by a tensor product of the single walker
time evolution operators (\ref{U1}). We describe the state of the
system by vectors
\begin{equation}
\psi(m,n,t)= \left(%
\begin{array}{c}
  \psi_{LL}(m,n,t) \\
  \psi_{RL}(m,n,t) \\
  \psi_{LR}(m,n,t) \\
  \psi_{RR}(m,n,t) \\
\end{array}%
\right),
\end{equation}
where e.g. the component $\psi_{RL}(m,n,t)$ is the amplitude of
the state where the first walker is on $m$ with the internal state
$|R\rangle$ and the second walker is on $n$ with the internal
state $|L\rangle$. The state of the two walkers at time $t$ is
then given by
\begin{equation}\label{dist}
|\psi(t)\rangle = \sum_{m,n}\sum_{i,j=\
R,L}\psi_{ij}(m,n,t)|m,i\rangle_1|n,j\rangle_2.
\end{equation}
The conditional probability that the first walker is on a site $m$
at time $t$, provided that the second walker is at the same time
at site $n$, is defined by
\begin{equation}\label{P2}
P(m,n,t) = \sum_{i,j=L,R}|\langle m,i|\langle
n,j|\psi(t)\rangle|^2 = \sum_{i,j=L,R}|\psi_{ij}(m,n,t)|^2.
\end{equation}
Note that if we would consider a single random walker but on a two
dimensional lattice, with two independent Hadamard coins for each
spatial dimension, (\ref{P2}) will give the probability distribution
generated by such a two dimensional walk. This shows the relation
between a one dimensional walk with two walkers and a two
dimensional walk.

The reduced probabilities for the first and the second walker are
given by
\begin{equation}
P_1(m,t) = \sum_n P(m,n,t),\quad P_2(n,t) = \sum_m P(m,n,t).
\end{equation}
We can rewrite them with the help of the reduced density operators
of the corresponding walkers
\begin{equation}
\rho_i(t) = Tr_{j\neq i}|\psi(t)\rangle\langle\psi(t)|,
\end{equation}
in the form
\begin{equation}
P_i(m,t) = \sum_j \langle m,j|\rho_i(t)|m,j\rangle.
\end{equation}

The dynamics of the two walkers is determined by the single walker
motion. Since we can always decompose the initial state of the two
walkers into a linear combination of a tensor product of a single
walker states and because the time evolution is also given by a
tensor product of two unitary operators, the shape of the state will
remain unchanged. Thus we can fully describe the time evolution of
the two random walkers with the help of the single walker
wave-functions. A similar relation holds for the probability
distribution (\ref{P2}). Moreover, in the particular case when the
two walkers are initially in a factorized state
\begin{equation}\label{fact}
|\psi(0)\rangle = \left(
\sum_{m,i}\psi_{1i}(m,0)|m,i\rangle_1\right)\otimes\left(\sum_{n,j}\psi_{2j}(n,0)|n,j\rangle_2\right),
\end{equation}
which translates into
$\psi_{ij}(m,n,0)=\psi_{1i}(m,0)\psi_{2j}(n,0)$, the probability
distribution remains a product of single walker probability
distributions
\begin{eqnarray}\label{fp}
P(m,n,t) &=&
(|\psi_{1L}(m,t)|^2+|\psi_{1R}(m,t)|^2)(|\psi_{2L}(n,t)|^2+|\psi_{2R}(n,t)|^2) \nonumber\\
&=& P_1(m,t)P_2(n,t).
\end{eqnarray}
However, when the initial state of the two walkers is entangled
\begin{equation}\label{ent}
|\psi(0)\rangle = \sum_\alpha\left\{\left(
\sum_{m,i}\psi^\alpha_{1i}(m,0)|m,i\rangle_1\right)\otimes\left(\sum_{n,j}\psi^\alpha_{2j}(n,0)|n,j\rangle_2\right)\right\},
\end{equation}
the probability distribution cannot be expressed in terms of single
walker distributions but probability amplitudes
\begin{equation}\label{ment}
P(m,n,t) = \sum_{i,j=L,R} \left|
\sum_\alpha\psi^\alpha_{1i}(m,t)\psi^\alpha_{2j}(n,t)\right|^2.
\end{equation}

As we have shown above, when the two walkers are initially in a
factorized state, the probability distribution (\ref{fp}) is a
product of the reduced distributions and thus the positions of the
walkers are independent. On the other hand, when they are
entangled, the probability distribution differs from a product of
the reduced probabilities and the positions of the walkers are
correlated. However, notice that the correlations are present also
in the classical random walk with two walkers, if we consider
initial conditions of the following form
\begin{equation}\label{clcor}
P(m,n,0) = \sum_\alpha P_1^\alpha(m,0)P_2^\alpha(n,0).
\end{equation}
The difference between (\ref{ment}) and (\ref{clcor}) is that in the
quantum case we have probability amplitudes not probabilities. The
effect of the quantum mechanical dynamics is the interference in
(\ref{ment}).

Let us now define the meeting problem. We ask for the probability
that the two walkers will be detected at the position $m$ after $t$
steps. This probability is given by the norm of the vector
$\psi(m,m,t)$
\begin{equation}\label{md}
M_D(m,t) = \sum_{i,j=L,R}|\psi_{ij}(m,m,t)|^2 =  P(m,m,t).
\end{equation}
As we have seen above for a particular case when the two walkers are
initially in a factorized state of the form (\ref{fact}) this can be
further simplified to the multiple of the probabilities that the
individual walkers will reach the site. However, this not possible
in the situation when the walkers are initially entangled
(\ref{ent}). The entanglement introduced in the initial state of the
walkers leads to the correlations between the walkers position and
thus the meeting probability is no longer a product of the single
walker probabilities.

\subsection{Indistinguishable walkers}

We now analyze the situation when the two walkers are
indistinguishable. Because we work with indistinguishable particles
we use the Fock space and creation operators, we use symbols
$a_{(m,i)}^\dagger$ for bosons and $b_{(n,j)}^\dagger$ for fermions,
e.g. $a_{(m,i)}^\dagger$ creates one bosonic particle at position
$m$ with the internal state $|i\rangle$. The time evolution is now
given by the transformation of the creation operators, e.g. for
bosons a single displacement is described by the following
transformation of the creation operators
\begin{eqnarray}
\nonumber a_{(m,L)}^\dagger & \longrightarrow & \frac{1}{\sqrt{2}}\left(a_{(m-1,L)}^\dagger+a_{(m+1,R)}^\dagger\right)\\
a_{(m,R)}^\dagger & \longrightarrow &
\frac{1}{\sqrt{2}}\left(a_{(m-1,L)}^\dagger-a_{(m+1,R)}^\dagger\right),
\end{eqnarray}
similarly for the fermionic walkers. The dynamics is defined on a
one-particle level. We will describe the state of the system by
the same vectors $\psi(m,n,t)$ as for the distinguishable walkers.
The state of the two bosonic and fermionic walkers analogous to
(\ref{dist}) for the distinguishable is given by
\begin{eqnarray}
\nonumber |\psi_B(t)\rangle & = & \sum_{m,n}\sum_{i,j=
L,R}\frac{1}{2}(\psi_{ij}(m,n,t)+\psi_{ji}(n,m,t))a_{(m,i)}^\dagger
a_{(n,j)}^\dagger|vac\rangle,\\
|\psi_F(t)\rangle & = & \sum_{m,n}\sum_{i,j=
L,R}\frac{1}{2}(\psi_{ij}(m,n,t)-\psi_{ji}(n,m,t))b_{(m,i)}^\dagger
b_{(n,j)}^\dagger|vac\rangle,
\end{eqnarray}
where $|vac\rangle$ is the vacuum state. In the case of two
bosonic walkers initially in the same state we have to include an
additional factor of $1/\sqrt{2}$, to ensure proper normalization.

The conditional probability distribution is given by
\begin{eqnarray}
P_{B,F}(m,n,t) &=& \sum_{i,j=L,R}|\langle
1_{(m,i)}1_{(n,j)}|\psi_{B,F}(t)\rangle|^2 \nonumber \\
&=& \sum_{i,j = L,R}|\psi_{ij}(m,n,t)\pm\psi_{ji}(n,m,t)|^2,
\end{eqnarray}
for $m\neq n$, and for $m=n$
\begin{eqnarray}\label{mbf}
\nonumber P_B(m,m,t) & = & |\langle 2_{(m,L)}|\psi_B(t)\rangle|^2+|\langle 2_{(m,R)}|\psi_B(t)\rangle|^2+ \\
\nonumber && \quad+ |\langle 1_{(m,L)}1_{(m,R)}|\psi_B(t)\rangle|^2 \\
\nonumber & = & 2|\psi_{LL}(m,m,t)|^2+2|\psi_{RR}(m,m,t)|^2+\\
\nonumber && \quad +|\psi_{LR}(m,m,t)+\psi_{RL}(m,m,t)|^2\\
\nonumber & = & M_B(m,t),\\
\nonumber P_F(m,m,t) & = & |\langle 1_{(m,L)}1_{(m,R)}|\psi_F(t)\rangle|^2  \\
\nonumber  &=& |\psi_{LR}(m,m,t)-\psi_{RL}(m,m,t)|^2\\
 & = & M_F(m,t).
\end{eqnarray}
The diagonal terms of the probability distribution (\ref{mbf})
define the meeting probability we wish to analyze.

Let us now specify the meeting probability for the case when the
probability amplitudes can be written in a factorized form
$\psi_{ij}(m,n,t) = \psi_{1i}(m,t)\psi_{2j}(n,t)$, which for the
distinguishable walkers corresponds to the situation when they are
initially not correlated. In this case the meeting probabilities are
given by
\begin{eqnarray}\label{mb}
M_B(m,t) =
2|\psi_{1L}(m,t)\psi_{2L}(m,t)|^2+2|\psi_{1R}(m,t)\psi_{2R}(m,t)|^2 \nonumber \\
+|\psi_{1L}(m,t)\psi_{2R}(m,t)+\psi_{1R}(m,t)\psi_{2L}(m,t)|^2,
\end{eqnarray}
for bosons and
\begin{equation}\label{mf}
M_F(m,t) =
|\psi_{1L}(m,t)\psi_{2R}(m,t)-\psi_{1R}(m,t)\psi_{2L}(m,t)|^2,
\end{equation}
for fermions. We see that they differ from the formulas for the
distinguishable walkers, except a particular case when the two
bosons start in the same state, i. e. $\psi_1(m,0) = \psi_2(m,0) =
\psi(m,0) $ for all integers $m$. For this initial state we get
\begin{eqnarray}
M_B(m,t) &=&
|\psi_{L}(m,t)|^4+|\psi_{R}(m,t)|^4+2|\psi_{L}(m,t)\psi_{R}(m,t)|^2 \nonumber \\
&=& (|\psi_L(m,t)|^2+|\psi_R(m,t)|^2)^2 \nonumber \\
 &=& P^2(m,t) ,
\end{eqnarray}
which is the same as for the case of distinguishable walkers
starting at the same point with the same internal state.

We conclude this section by emphasizing that except the case of
two bosons with the same initial state the probability of meeting
differs from the case of distinguishable particles and we have to
use the probability amplitudes, whereas for the distinguishable
walkers we can reduce it to one-particle probabilities.

\section{Analysis of the meeting problem}
Let us now compare the meeting problem in the classical and quantum
case. We will study the two following two probabilities: the total
probability of meeting after $t$ step have been performed defined by
\begin{equation}\label{2}
M(t) = \sum_{m}M(m,t) ,
\end{equation}
and the overall probability of meeting during some period of steps
$T$ defined as
\begin{equation}\label{ov}
\overline{M}(T) = 1-\prod_{t=1}^T\left(1-M(t)\right) .
\end{equation}
The total probability of meeting $M(t)$ is the probability that the
two walkers meet at time $t$ anywhere on the lattice, the overall
probability of meeting $\overline{M}(T)$ is the probability that
they meet at least once anywhere on the lattice during the first $T$
steps.

\subsection{Distinguishable walkers}

We first concentrate on the influence of the initial state on the
meeting probability for the distinguishable walkers. We consider
three situation, the walker will start localized with some initial
distance $2d$ (for odd initial distance they can never meet, without
loss of generality we assume that the first starts at position zero
and the second at position $2d$), with the coin states: first, right
for the first walker and left for the second
\begin{equation}
\psi_{RL}(0,2d,0) = 1 ,
\end{equation}
second, the symmetric initial conditions
$1/\sqrt{2}(|L\rangle+i|R\rangle)$ for both
\begin{equation}
\psi(0,2d,0) = \frac{1}{2}\left( \begin{array}{c}
  1 \\
  i \\
  i \\
  -1
\end{array}\right) ,
\end{equation}
and third, left for the first walker and right for the second
\begin{equation}
\psi_{LR}(0,2d,0) = 1.
\end{equation}
In the first case the probability distributions of the walkers are
biased to the right for the first walker, respectively to the left
for the second, and thus the walkers are moving towards each other.
In the second case the walkers mean positions remain unchanged, as
for this initial condition the single walker probability
distribution is symmetric and unbiased. In the last case the walkers
are moving away from each other as their probability distributions
are biased to the left for the first one and to the right for the
second.

Let us now specify the meeting probabilities. Recalling the
expressions for the probability distributions $P^L$ and $P^R$, we can
write the total meeting probabilities with the help of the relations
(\ref{fp}), (\ref{md}) and (\ref{2}) as
\begin{eqnarray}\label{mq}
\nonumber M_{RL}(t,d) & = & \sum_m P^R(m,t)P^L(m-2d,t) \\
\nonumber M_{S}(t,d) & = &
\sum_m\frac{P^L(m,t)+P^R(m,t)}{2}\frac{P^L(m-2d,t)+P^L(m-2d,t)}{2}\\
M_{LR}(t,d) & = & \sum_mP^L(m,t)P^R(m-2d,t).
\end{eqnarray}
We see that the meeting probability is fully determined by the
single walker probability distribution.

The figure 2 shows the time evolution of the total probability of
meeting for the three studied situations and compares it with the
classical case. The initial distance is set to 10 and 20 points.
The plot clearly shows the difference between the quantum and the
classical case.
\begin{figure}
\begin{center}
\includegraphics[width=7cm]{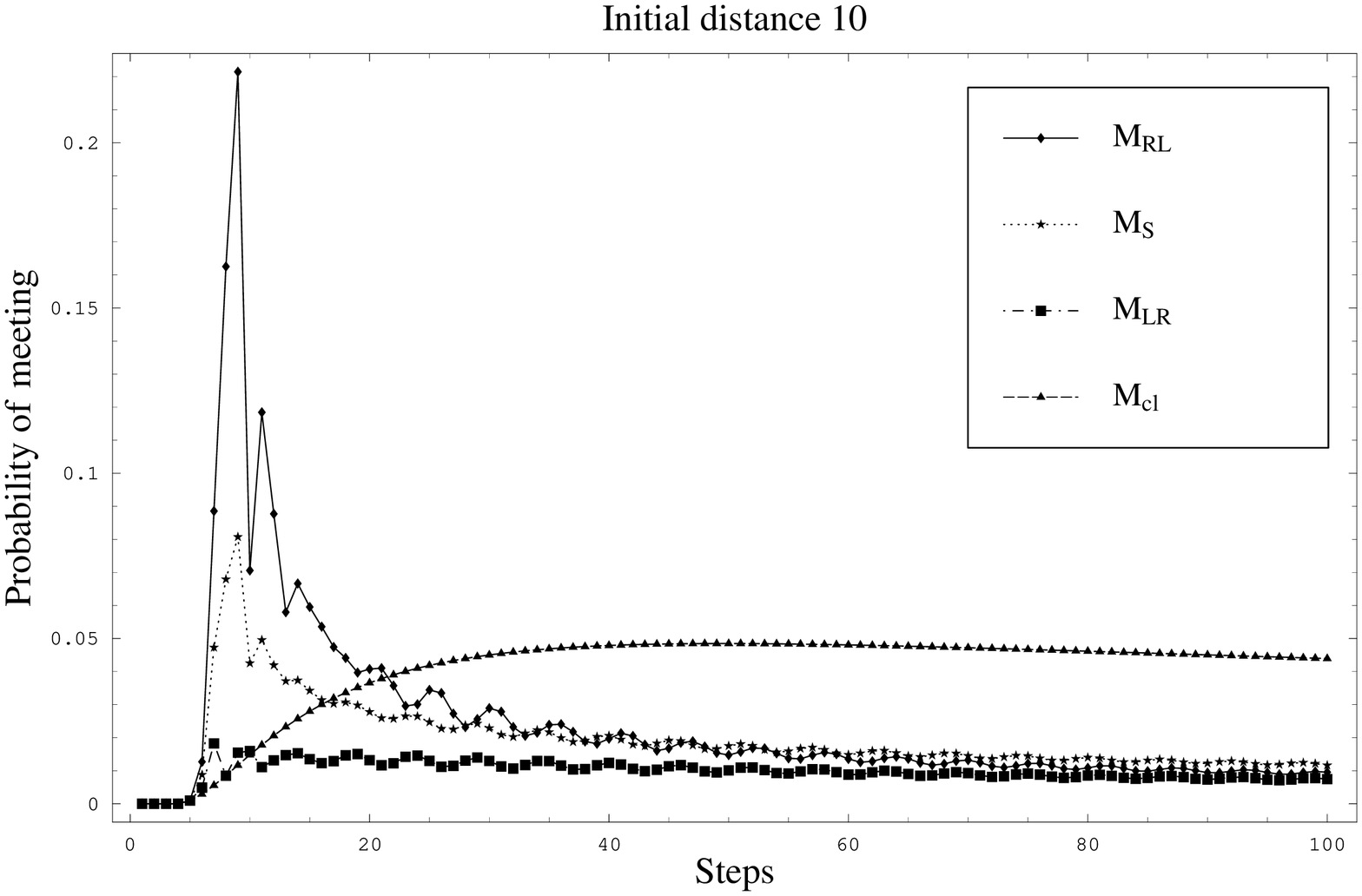}
\includegraphics[width=7cm]{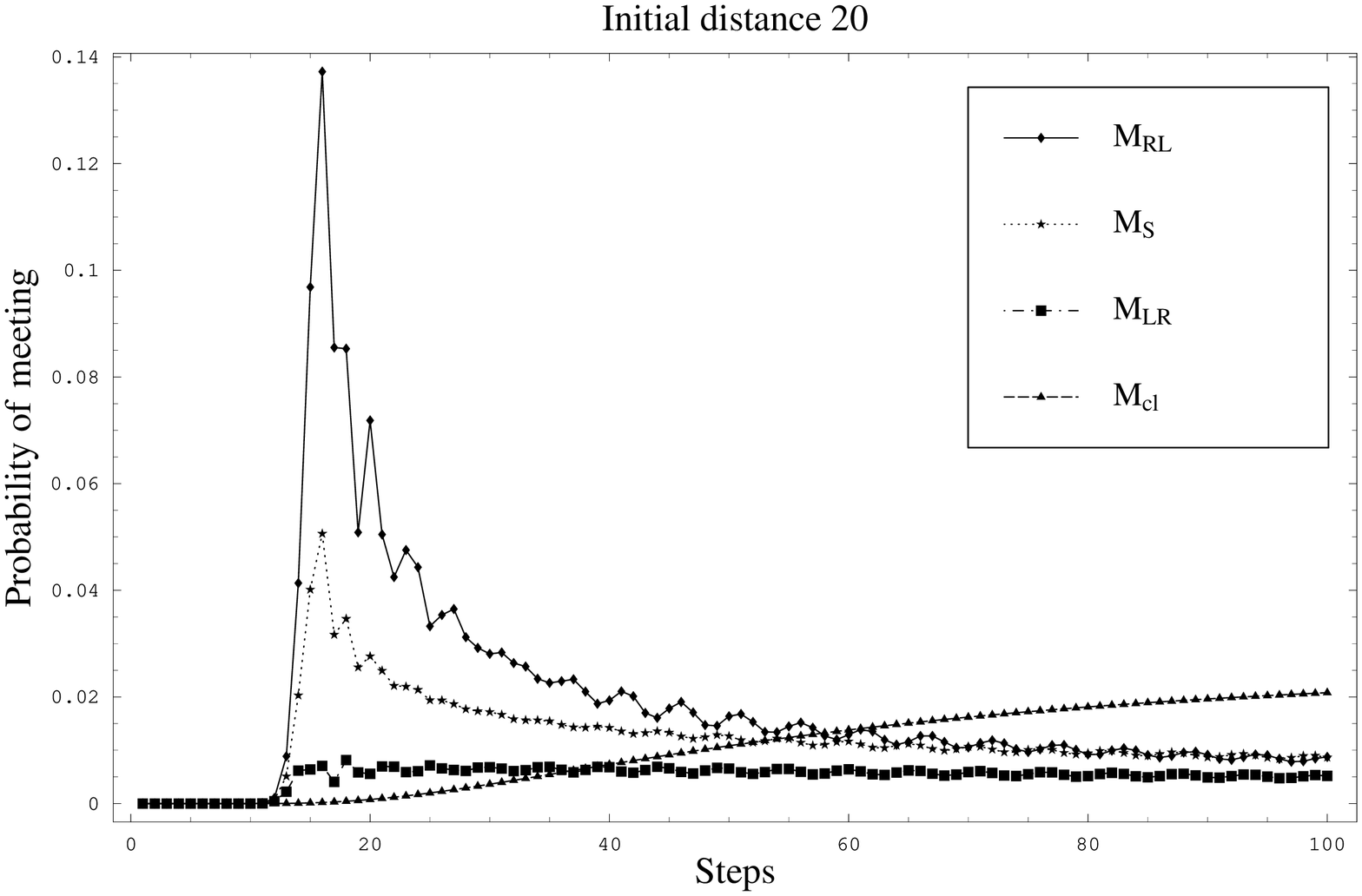}
\end{center}
\caption{Time evolution of the total probability of meeting for
the three types of initial states and the classical random walk
with two walkers. The initial distance is set to 10 and 20
points.}
\end{figure}

We derive some of the properties of the classical meeting problem in
the Appendix A. The main results are the following. The meeting
probability can be estimated by
\begin{equation}
M_{cl}(t,d)\approx\frac{1}{\sqrt{\pi t}} \exp(-\frac{d^2}{t}).
\end{equation}
This function has a maximum for $t=2d^2$, the peak value is given
approximately by
\begin{equation}\label{mclmax}
M_{cl}(2d^2,d)\approx\frac{1}{\sqrt{2\pi e}d}.
\end{equation}

In contrast to the classical walk, in the quantum case the meeting
probability is oscillatory. The oscillations come from the
oscillations of the single walker probability distribution. After
some rapid oscillations in the beginning we get a periodic function
with the characteristic period of about six steps, independent of
the initial state. In the quantum case the maximum of the meeting
probability is reached sooner than in the classical case - the
number of steps needed to hit the maximum is linear in the initial
distance $d$. This can be understood from the shape of the walkers
probability distribution. The maximum of the meeting probability is
obtained when the peaks of the probability distribution of the first
and second walker overlap. If the initial distance between the two
walkers is $2d$ then the peaks will overlap approximately after
$\sqrt{2}d$ steps. The value of the maximum depends on the choice of
the initial state.

Let us derive analytical formulas for the meeting probabilities in
the quantum case. For $t\geq\sqrt{2}d$ we consider the slowly
varying part of the walker probability distribution (\ref{pslow}),
(\ref{pslowsym}) and estimate the sums in (\ref{mq}) by the
integrals

\begin{eqnarray}\label{M1}
\nonumber M_{RL}(t,d) & \approx & \frac{2}{\pi^2 t^2}\int\limits_{2d-\frac{t}{\sqrt{2}}}^{\frac{t}{\sqrt{2}}}\frac{dx}{(1-\frac{x}{t})(1+\frac{x-2d}{t})\sqrt{1-2\frac{x^2}{t^2}}\sqrt{1-2\frac{(x-2d)^2}{t^2}}}\\\nonumber\\
\nonumber M_{S}(t,d) & \approx & \frac{2}{\pi^2 t^2}\int\limits_{2d-\frac{t}{\sqrt{2}}}^{\frac{t}{\sqrt{2}}}\frac{dx}{(1-\frac{x^2}{t^2})(1-\frac{(x-2d)^2}{t^2})\sqrt{1-2\frac{x^2}{t^2}}\sqrt{1-2\frac{(x-2d)^2}{t^2}}}\\\nonumber\\
M_{LR}(t,d) & \approx & \frac{2}{\pi^2
t^2}\int\limits_{2d-\frac{t}{\sqrt{2}}}^{\frac{t}{\sqrt{2}}}\frac{dx}{(1+\frac{x}{t})(1-\frac{x-2d}{t})\sqrt{1-2\frac{x^2}{t^2}}\sqrt{1-2\frac{(x-2d)^2}{t^2}}}
\end{eqnarray}
which can be evaluated in terms of elliptic integrals. Notice that
the integrals diverge for $d=0$, i.e. for the case when the two
walkers start at the same point. We will discuss this particular
case later, for now we will suppose that $d>0$. The formulas
(\ref{M1}) can expressed in the form
\begin{eqnarray}\label{Mq}
\nonumber M_{RL}(t,d) & \approx & F_+\{2(t-d)(t-(4-2\sqrt{2})d)K(a)+\\
\nonumber & &
+\sqrt{2}((t-(4+2\sqrt{2})d)(t-(4-2\sqrt{2})d)\Pi(b_+|a)-t^2\Pi(c_+|a))\}\\\nonumber
\\ \nonumber M_{S}(t,d) & \approx & \frac{ \pi^2 F_+F_-}{4}\{
16d(t^2-d^2)(t+(4+2\sqrt{2})d)(t-(4-2\sqrt{2})d)K(a)+\\ \nonumber
& &
+\sqrt{2}(t+(4+2\sqrt{2})d)(t-(4+2\sqrt{2})d)(t+(4-2\sqrt{2})d)\times \\ \nonumber & & \times(t-(4-2\sqrt{2})d)((t+d)\Pi(b_+|a)+(t-d)\Pi(b_-|a))-\\
\nonumber & &
-\sqrt{2}t^2((t+d)(t+(4+2\sqrt{2})d)(t+(4-2\sqrt{2})d)\Pi(c_+|a)+\\
\nonumber & &
+(t-d)(t-(4+2\sqrt{2})d)(t-(4-2\sqrt{2})d)\Pi(c_-|a))\}\\\nonumber
\\ \nonumber M_{LR}(t,d) & \approx &
F_-\{2(t+d)(t+(4+2\sqrt{2})d)K(a)-\\ 
\nonumber & & -\sqrt{2}((t+(4+2\sqrt{2})d)(t+(4-2\sqrt{2})d)\Pi(b_-|a)- \\
&& -t^2\Pi(c_-|a))\}
\end{eqnarray}
where $K(a)$ is the complete elliptic integral of the first kind
and $\Pi(x|a),\Pi(y|a)$ are the complete elliptic integrals of the
third kind (see e.g. \cite{abramowitzstegun}, chapter 17),
$a,b_\pm,c_\pm$ and $F_\pm$ are given by
\begin{eqnarray}
\nonumber F_\pm & = & \frac{2t}{\pi^2 d(t\mp
d)(t(2+\sqrt{2})\mp 4d)(t(2-\sqrt{2})\mp 4d))} \\
\nonumber a & = & i\sqrt{\frac{t^2}{2d^2}-1} \\
\nonumber b_\pm & = & \frac{(1\pm \sqrt{2})(t-\sqrt{2}d)}{d(\sqrt{2}\mp 2)}\\
c_\pm & = & \frac{(t(\sqrt{2}\mp
2)+4d)(t-\sqrt{2}d)}{\sqrt{2}d(t(\sqrt{2}\pm 2)-4d)}.
\end{eqnarray}
From the relations (\ref{Mq}) we can estimate the peak value of
the meeting probability
\begin{eqnarray}\label{Mmax}
\nonumber M_{RL}(\sqrt{2}d,d) & \approx & \frac{2-3\sqrt{2}}{\pi
d(18-13\sqrt{2})}\\ \nonumber M_{S}(\sqrt{2}d,d) & \approx &
\frac{2}{\pi d}\\ M_{LR}(\sqrt{2}d,d) & \approx &
\frac{2+3\sqrt{2}}{\pi d(18+13\sqrt{2})}.
\end{eqnarray}
The peak value shows a $1/d$ dependence on the initial distance
between the two walker in all three studied cases, similar to the
classical situation (\ref{mclmax}).

Let us now analyze the asymptotic behavior of the meeting
probability. In the Appendix A we show that in the classical case
the meeting probability can be estimated by
\begin{equation}\label{apcl}
M_{cl}(t,d)\approx\frac{1}{\sqrt{\pi t}}(1-\frac{d^2}{t}).
\end{equation}
In the quantum case we begin with the observation that the
coefficients at the highest power of $t$ with the elliptic integrals
of the third kind are the same but with the opposite sign for
$\Pi(b|a)$ and $\Pi(c|a)$. Moreover, $b_\pm$ and $c_\pm$ go like
$-t$ as $t$ approaches infinity, and thus all of the $\Pi$ functions
have the same asymptotic behavior. Due to the opposite sign for
$\Pi(b|a)$ and $\Pi(c|a)$ the leading order terms cancel and the
contribution from this part to the meeting probability is of higher
order of $1/t$ compared to the contribution from the complete
elliptic integral of the first kind $K(a)$. The asymptotic of the
function $K(a)$ is given by
\begin{equation}
K(a)\approx
\frac{d\sqrt{2}\ln{\left(\frac{2\sqrt{2}t}{d}\right)}}{t}.
\end{equation}
Inserting this approximation into (\ref{Mq}) we find that the
leading order term for the meeting probability in all three studied
situations is given by (up to the constant factor $a$)
\begin{equation}\label{apq}
M_{D}(t,d) \approx a
\frac{\ln{\left(\frac{2\sqrt{2}t}{d}\right)}}{t}.
\end{equation}
Therefore, the meeting probability decays faster in the quantum case
and goes like $\ln(t/d)/t$ compared to the classical case
(\ref{apcl}). However, the decay is not quadratically faster, as one
could expect from the fact that the single walker probability
distribution spreads quadratically faster in the quantum walk. The
exponential peaks in the probability distribution of the quantum
random walker slow down the decay.

The above derived results holds for $d>0$, i.e. the initial distance
has to be non-zero. As we have mentioned before, for $d=0$ the
integrals (\ref{M1}) diverge, and therefore we cannot use this
approach for the estimation of the meeting probability. There does
not seem to be an easy analytic approach to the problem. However,
from the numerical results, the estimation
\begin{equation}
M_D(t) \approx b \frac{\ln{t}}{t}
\end{equation}
fits the data best ($b$ being a constant prefactor).

For illustration we plot in figure 3 the meeting probability and
the estimations on a long-time scale. In the first plot is the
case $M_{RL}$ with the initial distance 20 points, on the second
plot we have $M_{S}$ and the initial distance is zero.
\begin{figure}
\begin{center}
\includegraphics[width=7cm]{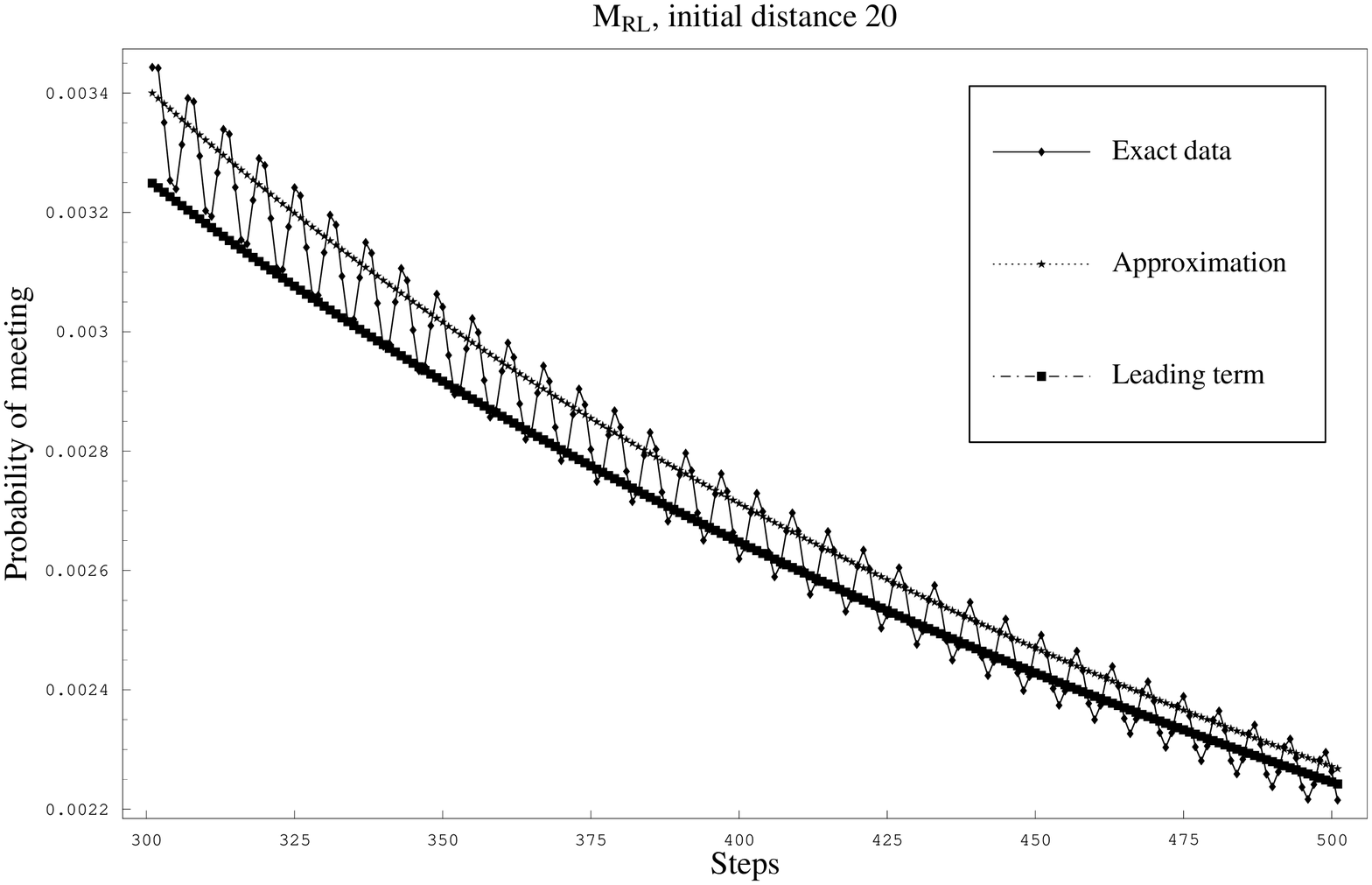}
\includegraphics[width=7cm]{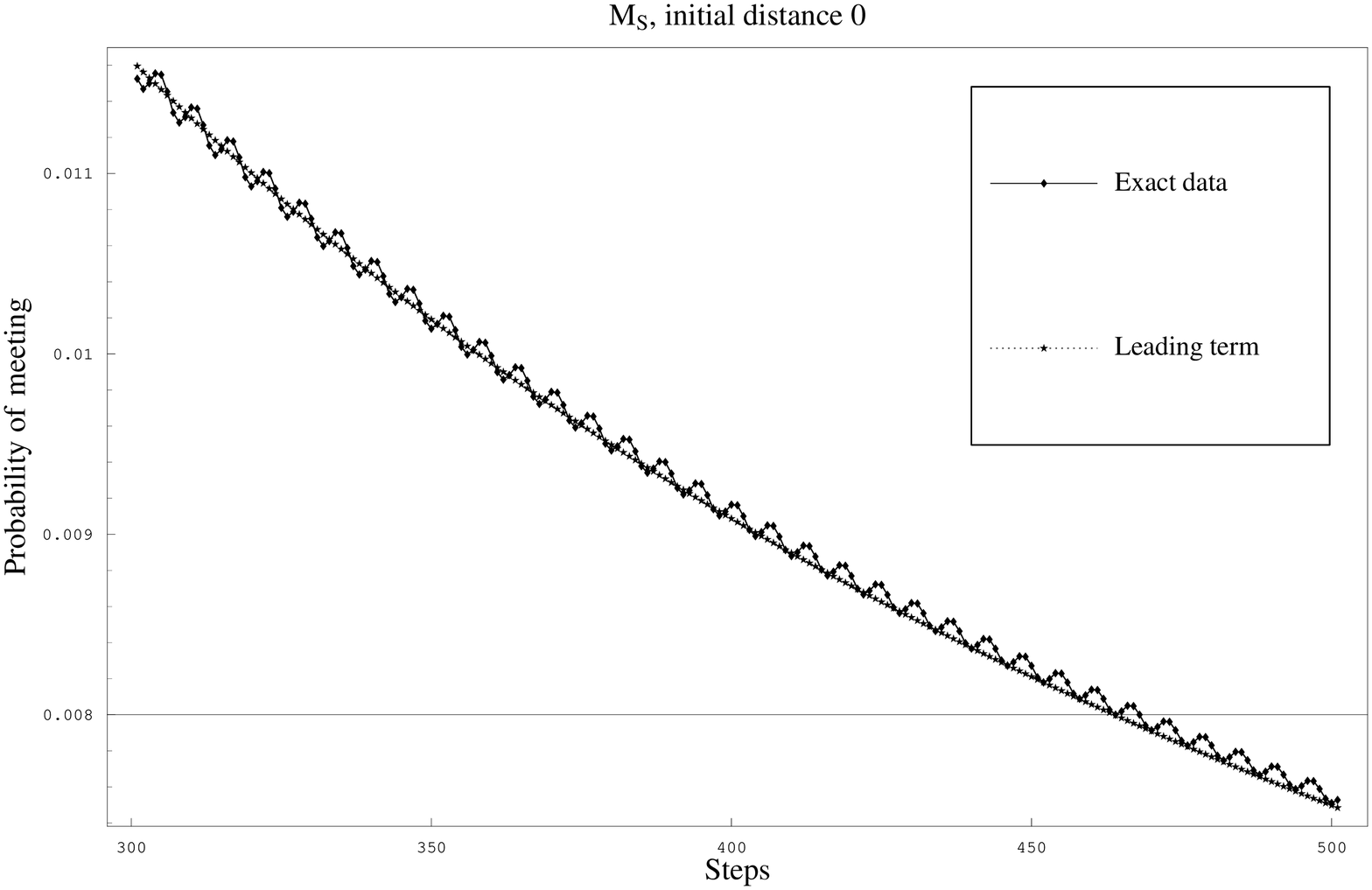}
\end{center}
\caption{Left: Comparison of the meeting probability with the
elliptic function estimation and the leading order term. The initial
distance is set to 20 points. Right: Comparison of the meeting
probability for the symmetric initial condition and zero initial
distance with the estimation.}
\end{figure}

Let us now focus on the overall meeting probability defined by
(\ref{ov}). In figure 4 we plot the overall probability that the two
walkers will meet during the first $T=100$ steps.
\begin{figure}
\begin{center}
\includegraphics[width=7cm]{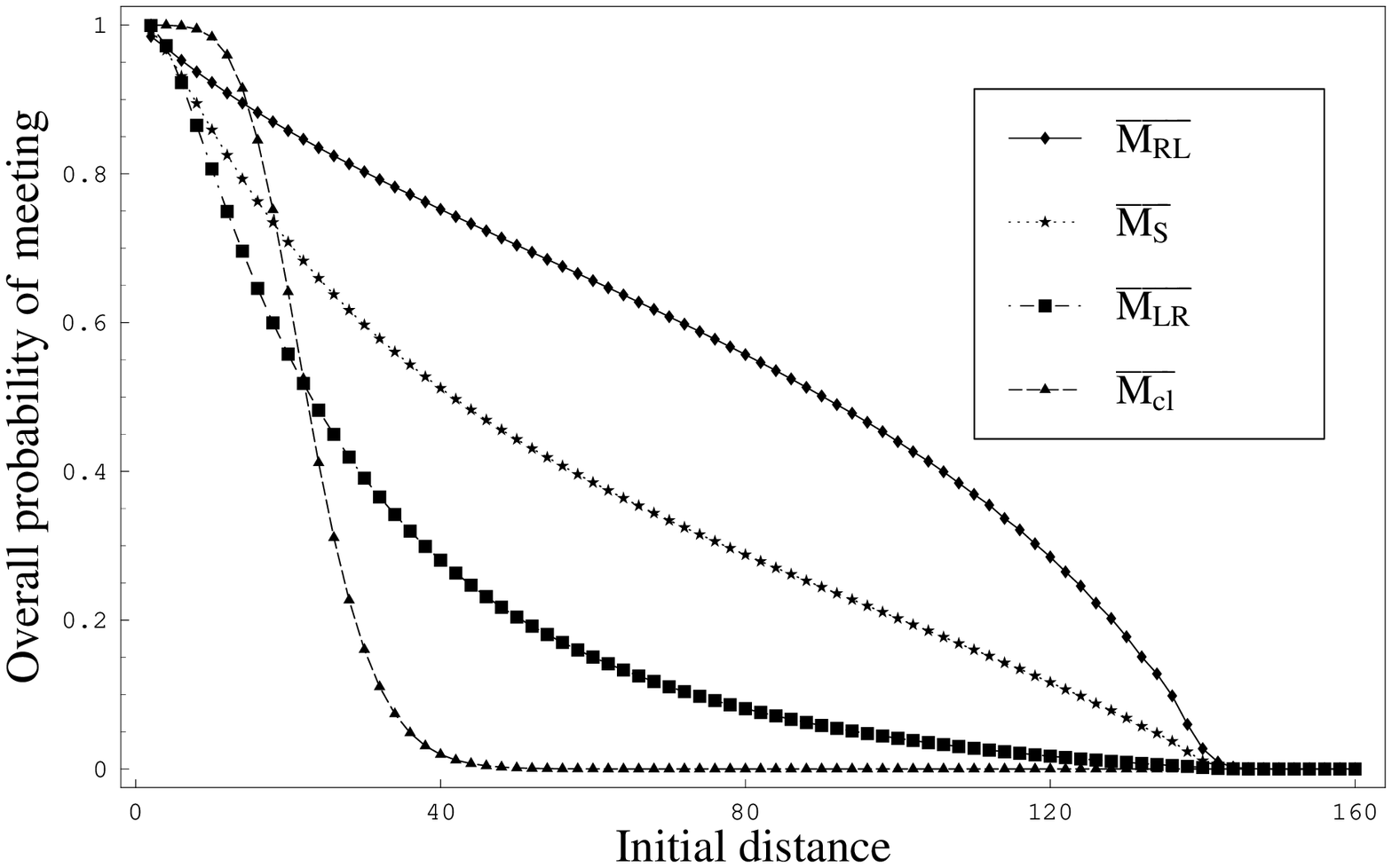}
\includegraphics[width=7cm]{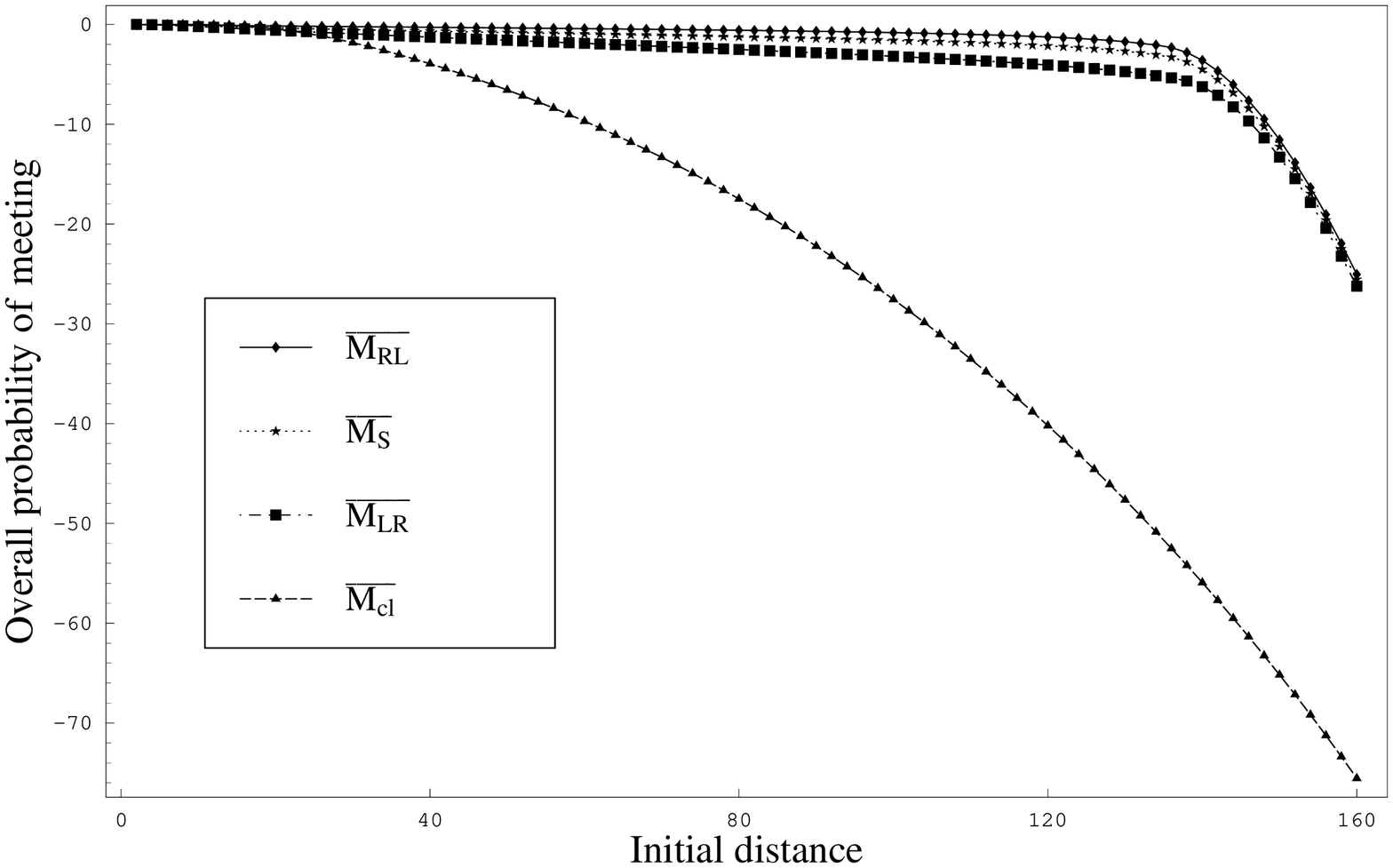}
\end{center}
\caption{The overall probability of meeting for two
distinguishable quantum and classical walker during first 100
steps as a function of the initial distance. Same plot on the log
scale. Only the values for even points are plotted since for odd
initial distance the walkers never meet.}
\end{figure}
On the first plot we present the difference between the three
studied quantum situations, whereas the second plot, where the
meeting probability is on the log scale, uncovers the difference
between the quantum and classical random walk. In the log scale plot
we can see that the overall meeting probability decays slower in the
quantum case then in the classical case, up to to the initial
distance of $\sqrt{2}T$. This can be understood by the shape and the
time evolution of a single walker probability distribution. After
$t$ steps the maximums of the probability distribution are around
the point $s\pm\sqrt{2}t$, where $s$ is the initial starting point
of the random walker. For $t=100$ steps the peaks are around the
points $s\pm 140$. So when the two walkers are initially more then
140 points away, the peaks do not overlap, and the probability of
meeting is given by just the tails of the single walker
distributions, which have almost classical behavior. From the first
plot we see that the overall meeting probability is broader in the
quantum case compared to the classical, which drops down very fast.
The numerical results in figure 5 show that the width of the overall
meeting probability grows linearly with the upper bound T, the slope
depends on the choice of the initial coin state. On the other hand
in the classical case the width grows like $\sqrt{T}$.
\begin{figure}
\begin{center}
\includegraphics[width=13cm]{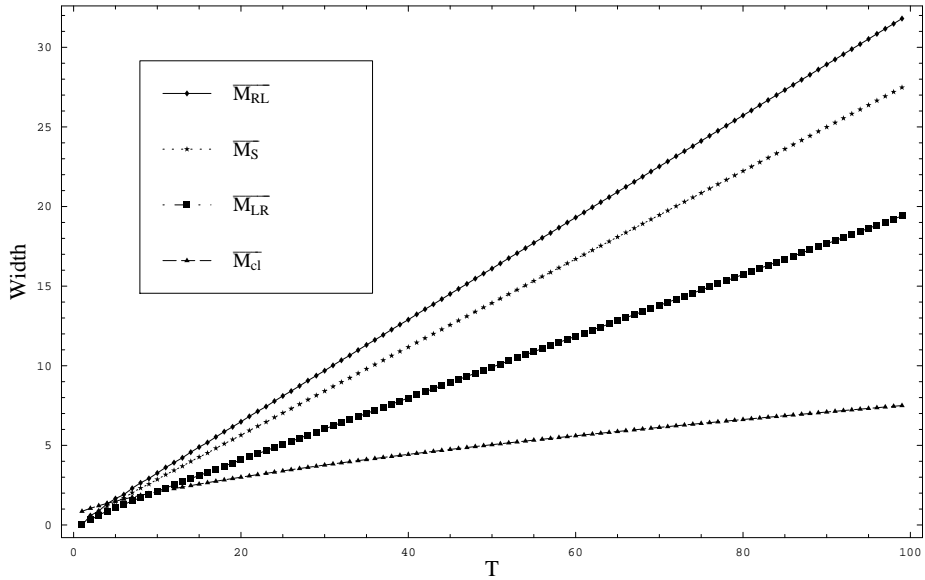}
\caption{Width of the overall meeting probability as a function of
the upper bound T.}
\end{center}
\end{figure}

Let us now analyze the overall meeting probability on a long time
scale. In the Appendix A we show that in the classical case the
overall meeting probability is approximately given by
\begin{equation}\label{cloe}
\overline{M_{cl}}(T,d) \approx
1-\exp\left(-2\sqrt{\frac{T}{\pi}}\exp({-\frac{d^2}{T}})\right)\exp\left(2d\textrm{Erfc}(\frac{d}{\sqrt{T}})\right).
\end{equation}
In the quantum case we consider $\ln{(1-\overline{M}_{D}(T,d))}$
($M_D$ stands for all three particular quantum cases) and estimate
it with the help of (\ref{Mq}) by
\begin{eqnarray}
\ln{(1-\overline{M}_{D}(T,d))} 
&=&\sum_{t=d}^T\ln{(1-M_{D}(t,d))}
\approx -\sum_{t=d}^T M_{D}(t,d)\nonumber \\
&\approx& -\int\limits_{\sqrt{2}d}^T M_{D}(t,d)dt.
\end{eqnarray}
Therefore we can estimate the overall meeting probability for
$T>\sqrt{2}d$ by
\begin{equation}\label{qoe}
\overline{M}_{D}(T,d) \approx
1-\exp\left(-\int\limits_{\sqrt{2}d}^T M_{D}(t,d)dt\right).
\end{equation}
The meeting probability in the quantum case (\ref{Mq}) involves
elliptic integrals in a rather complicated form. However, we can
estimate how fast the overall meeting probability converges to 1 for
a fixed initial distance. This is determined by the rate at which
the integral in (\ref{qoe}) diverges. Before we proceed notice that
\begin{equation}
\int\limits_{\sqrt{2}d}^{\alpha d}M_D(t,d)dt = C(\alpha),
\end{equation}
i.e. the integral does not depend on the initial distance $d$. For
large $t$ we can estimate the meeting probability by (\ref{apq})
and thus we can divide the integral in (\ref{qoe}) into
\begin{eqnarray}
\nonumber \int\limits_{\sqrt{2}d}^{T}M_D(t,d)dt & \approx &
\int\limits_{\sqrt{2}d}^{\alpha
d}M_D(t,d)dt+\frac{2\sqrt{2}}{\pi^2}\int\limits_{\alpha d}^{T}\frac{\ln{(\frac{2\sqrt{t}}{d})}}{t}dt\\
& \approx &
C(\alpha)+\frac{\sqrt{2}}{\pi^2}\ln^2{\left(\frac{2\sqrt{2}T}{d}\right)},
\end{eqnarray}
for appropriately large $\alpha$. Therefore the exponent in
(\ref{qoe}) goes like $-\ln^2(T)$ for large $T$. On the other hand,
in the classical case from the estimation (\ref{cloe}) we obtain
that the asymptotic behavior of the exponent is given $-\sqrt{T}$.
Comparing these two we conclude that the overall meeting probability
converges faster to one in the classical case.

To conclude, the quadratic speed-up of the width shown in figure 5
follows from the the quadratically faster spreading of the quantum
random walk. On the other hand for large times the meeting
probability decays faster in the quantum walk, which leads to the
slower convergence of the overall meeting probability.

\subsection{Effect of the entanglement for distinguishable walkers}

We will now consider the case when the two distinguishable walkers
are initially entangled. According to (\ref{ment}) the meeting
probability is no longer given by the product of single walker
probability distributions. However, it can be described using single
walker probability amplitudes. We consider the initial state of the
following form
\begin{equation}
|\psi(0)\rangle = |0, 2d\rangle\otimes|\chi\rangle,
\end{equation}
where $|\chi\rangle$ is one of the Bell states
\begin{eqnarray}\label{bell}
\nonumber |\psi^\pm\rangle &=& 
\frac{1}{\sqrt{2}}\left(|LR\rangle\pm|RL\rangle\right),\\
|\phi^\pm\rangle & = &
\frac{1}{\sqrt{2}}\left(|LL\rangle\pm|RR\rangle\right).
\end{eqnarray}
Recalling the probability amplitudes $\psi^{L(R)}(m,t)$ we can
write the probability distri\-butions of the two random walkers
(\ref{P2}) in the form
\begin{eqnarray}\label{pent}
\nonumber P_{\psi^\pm}(m,n,t)  =  \ \\
\quad \frac{1}{2} \sum_{i,j=L,R}|\psi_i^L(m,t)\psi_j^R(n-2d,t)\pm
\psi_i^R(m,t)\psi_j^L(n-2d,t)|^2, \nonumber \\
P_{\phi^\pm}(m,n,t)  =  \  \nonumber \\
\quad \frac{1}{2}
\sum_{i,j=L,R}|\psi_i^L(m,t)\psi_j^L(n-2d,t)\pm\psi_i^R(m,t)\psi_j^R(n-2d,t)|^2.
\end{eqnarray}
The meeting probabilities are given by the sum of the diagonal
terms in (\ref{pent})
\begin{eqnarray}\label{ment2}
\nonumber M_{\psi^\pm}(t,d)=  \nonumber \\
 \frac{1}{2}
\sum_m\left\{\sum_{i,j=L,R}|\psi_i^L(m,t)\psi_j^R(m-2d,t)
\pm\psi_i^R(m,t)\psi_j^L(m-2d,t)|^2\right\},\\
M_{\phi^\pm}(t,d)  = \nonumber \\
  \frac{1}{2}\sum_m\left\{
\sum_{i,j=L,R}|\psi_i^L(m,t)\psi_j^L(m-2d,t)\pm\psi_i^R(m,t)\psi_j^R(m-2d,t)|^2\right\}.
\end{eqnarray}
The reduced density operators for both coins are maximally mixed
for all four bell states (\ref{bell}). From this fact follows that
the reduced density operators of the walkers are
\begin{eqnarray}
\nonumber \rho_1(t) & = &
\frac{1}{2}\left(|\psi^L(t)\rangle\langle\psi^L(t)|+|\psi^R(t)\rangle\langle\psi^R(t)|\right)\\
\rho_2(t) & = &
\frac{1}{2}\left(|\psi^L_d(t)\rangle\langle\psi^L_d(t)|+|\psi^R_d(t)\rangle\langle\psi^R_d(t)|\right),
\end{eqnarray}
where $|\psi^{L(R)}_d(t)\rangle$ are analogous to
$|\psi^{L(R)}(t)\rangle$ but shifted by $2d$, i.e.
\begin{eqnarray}
\nonumber &|\psi^{L(R)}(t)\rangle = 
\frac{1}{2}\sum_m\left(\psi^{L(R)}_L(m,t)|m,L\rangle+\psi^{L(R)}_R(m,t)|m,R\rangle\right)\\
&|\psi^{L(R)}_d(t)\rangle =   \ \nonumber \\
& \frac{1}{2}\sum_m\left(\psi^{L(R)}_L(m-2d,t)|m,L\rangle+\psi^{L(R)}_R(m-2d,t)|m,R\rangle\right).
\end{eqnarray}
The reduced probabilities are therefore
\begin{eqnarray}\label{red}
\nonumber P_1(m,t) & = & \frac{1}{2}(P^L(m,t)+P^R(m,t))\\
P_2(m,t) & = & P_1(m-2d,t),
\end{eqnarray}
which are symmetric and unbiased. Notice that the product of the
reduced probabilities (\ref{red}) gives the probability
distribution of a symmetric case studied in the previous section.
Therefore to catch the interference effect in the meeting problem
we compare the random walks with entangled coin states
(\ref{bell}) with the symmetric case $M_{S}$. The figure 6 shows
the meeting probabilities and the difference $M_\chi-M_{S}$, the
initial distance between the two walker was chosen to be 10
points.
\begin{figure}
\begin{center}
\includegraphics[width=7cm]{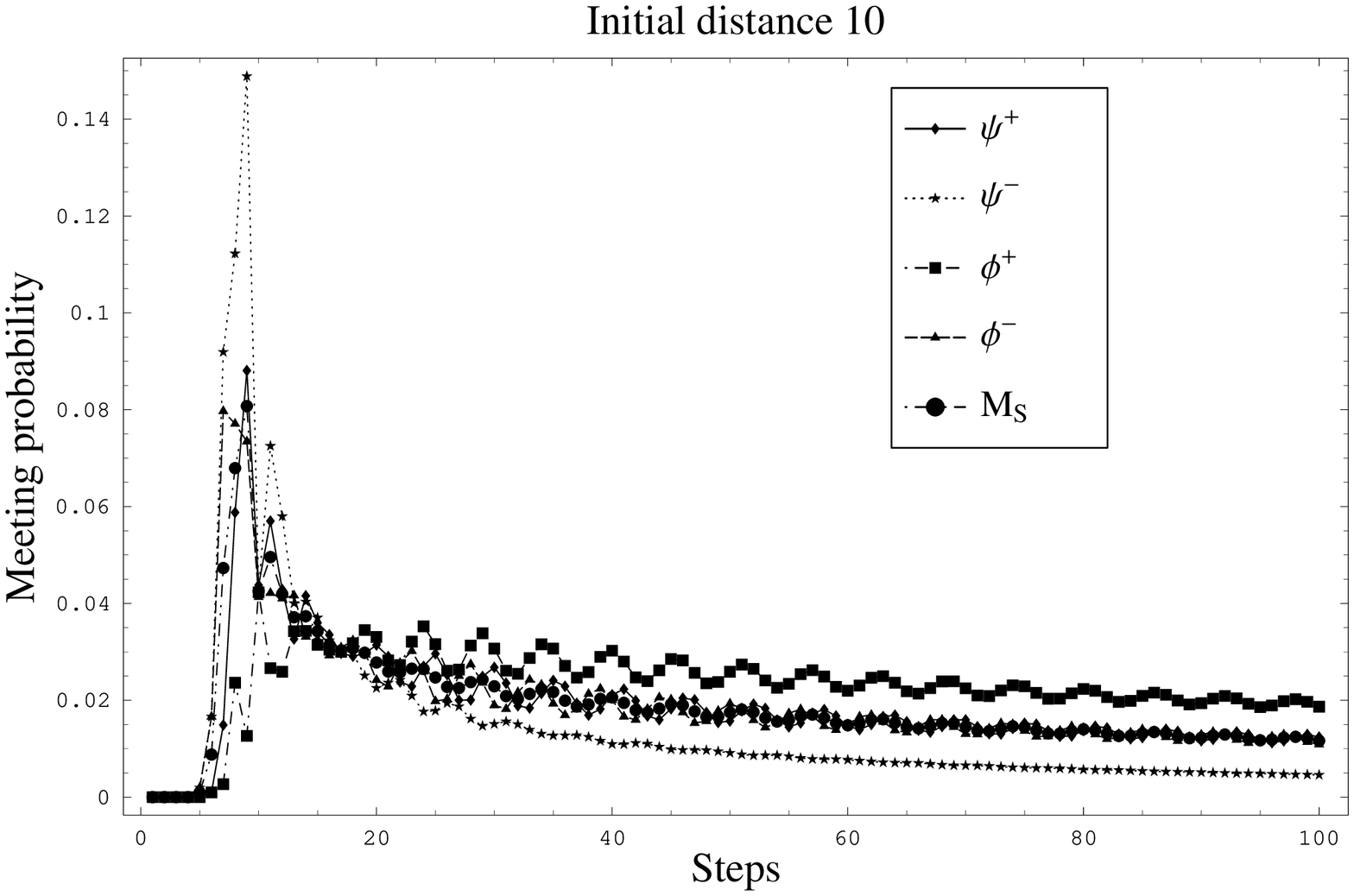}
\includegraphics[width=7cm]{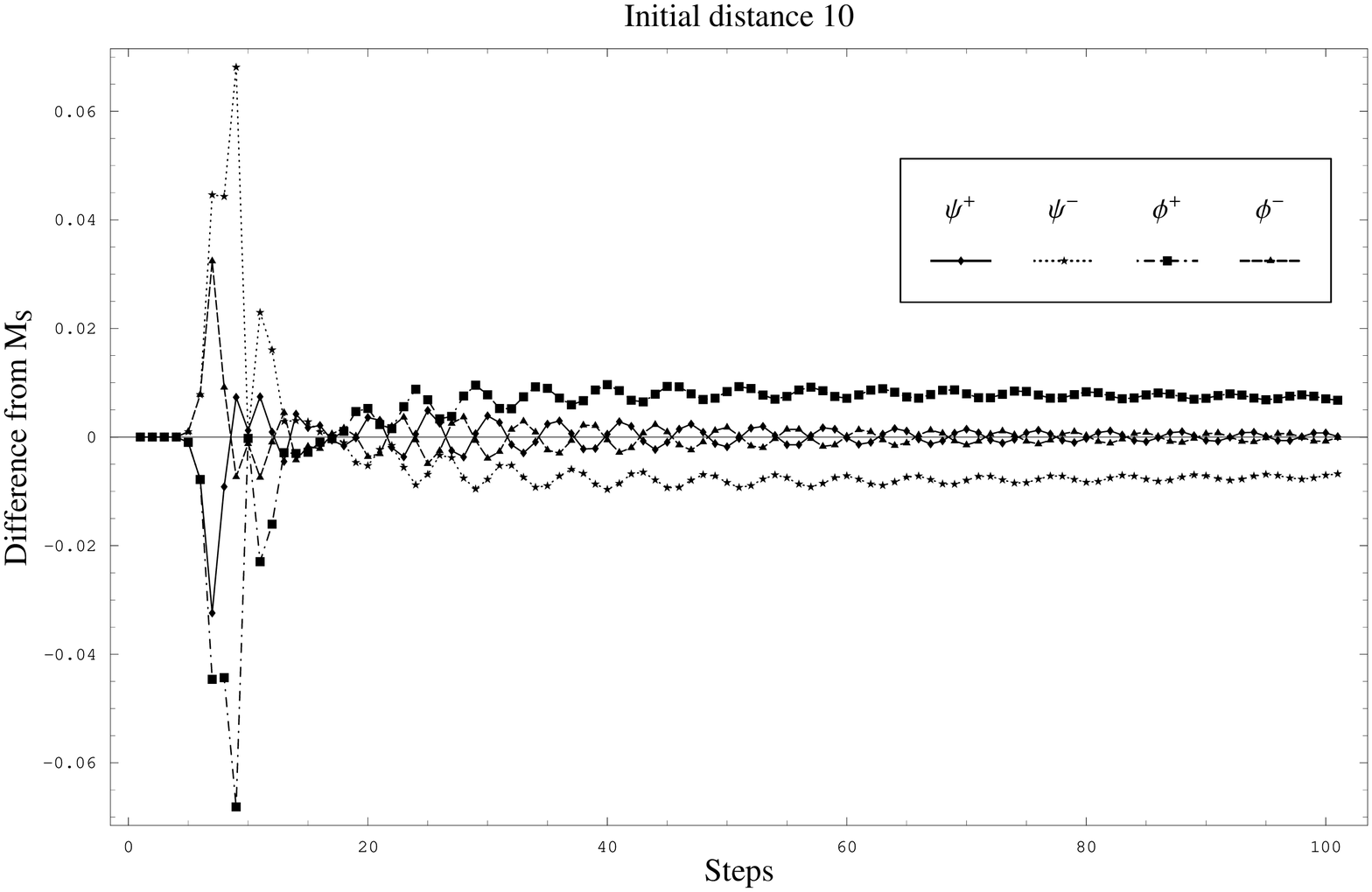}
\end{center}
\caption{Comparison of the meeting probability for the initially
entangled coins and the symmetric case. The difference in the
meeting probability.}
\end{figure}

We see that the effect of the entanglement could be both positive
or negative. Notice that
\begin{eqnarray}
\nonumber M_{\psi^-}(t,d)-M_S(t,d) & = &
-\left(M_{\phi^+}(t,d)-M_S(t,d)\right)\\
M_{\phi^-}(t,d)-M_S(t,d) & = &
-\left(M_{\psi^+}(t,d)-M_S(t,d)\right),
\end{eqnarray}
so the effect of $|\psi^-\rangle$ is opposite to $|\phi^+\rangle$
and $|\phi^-\rangle$ is opposite to $|\psi^+\rangle$. The main
difference is around the point $t\approx\sqrt{2}d$, i.e., the point
where for the factorized states the maximum of the meeting
probability is reached. The peak value is nearly doubled for
$M_{\psi^-}$ (note that for $M_S$ the peak value is given by
(\ref{Mmax}), which for $d=10$ gives $\approx 0.063$), but
significantly reduced for $M_{\phi^+}$. On the long time scale,
however, the meeting probability $M_{\psi^-}$ decays faster than in
other situations. According to the numerical simulations, the
meeting probabilities for $|\psi^+\rangle$ and $|\phi^\pm\rangle$
maintain the asymptotic behavior $\ln{t}/t$, but for
$|\psi^-\rangle$ it goes like
\begin{equation}
M_{\psi^-}(t,d)\sim\frac{1}{t}.
\end{equation}
The initial entanglement between the walkers influences the height
of the peaks giving the maximum meeting probability and affects also
the meeting probability on the long time scale.

\subsection{Indistinguishable walkers}

Let us briefly comment on the effect of the indistinguishability of
the walkers on the meeting probability. As an example, we consider
the initial state of the walkers of the form
$|1_{(0,R)}1_{(2d,L)}\rangle$, i.e. one walker starts at the site
zero with the right coin state and one starts at $2d$ with left
state. This corresponds to the case $M_{RL}$ for the distinguishable
walkers. The meeting probabilities are according to (\ref{mb}),
(\ref{mf}) given by
\begin{eqnarray}\label{bfmp}
\nonumber M_B(t,d) & = &
\sum_{m}\{2|\psi_L^R(m,t)|^2|\psi_L^L(m-2d,t)|^2+ \nonumber \\
&& +2|\psi_R^R(m,t)|^2|\psi_R^L(m-2d,t)|^2+\nonumber \\
&& 
+|\psi_L^R(m,t)\psi_R^L(m-2d,t)+\\
\nonumber & & +\psi_R^R(m,t)\psi_L^L(m-2d,t|^2\}\\
\nonumber
M_F(t,d) & = &
\sum_m\left(|\psi_L^R(m,t)\psi_R^L(m-2d,t)-\psi_R^R(m,t)\psi_L^L(m-2d,t)|^2\right).
\end{eqnarray}

In figure 7 we plot the meeting probabilities and the difference
$M_{B,F}-M_{RL}$.
\begin{figure}
\begin{center}
\includegraphics[width=7cm]{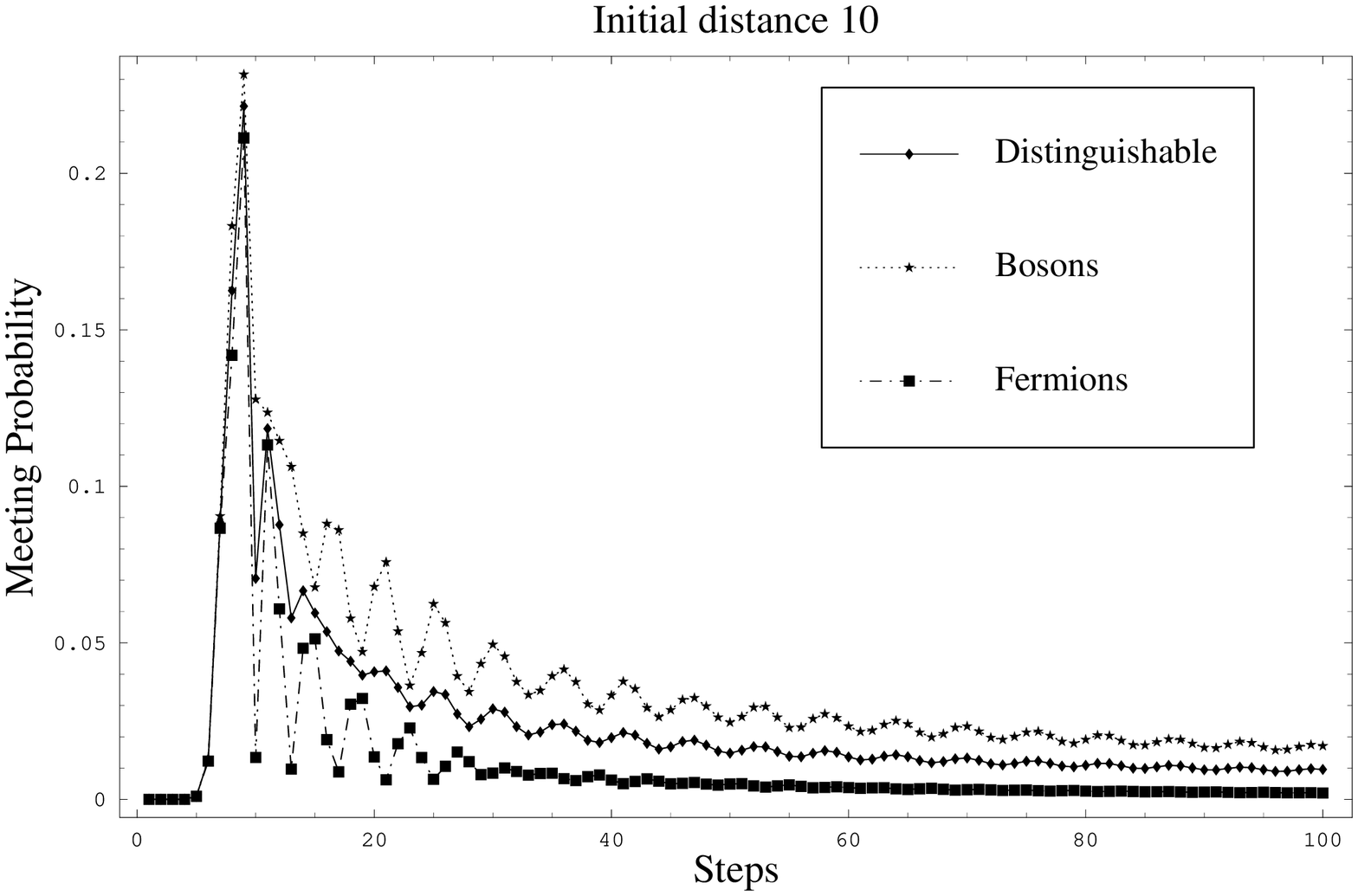}
\includegraphics[width=7cm]{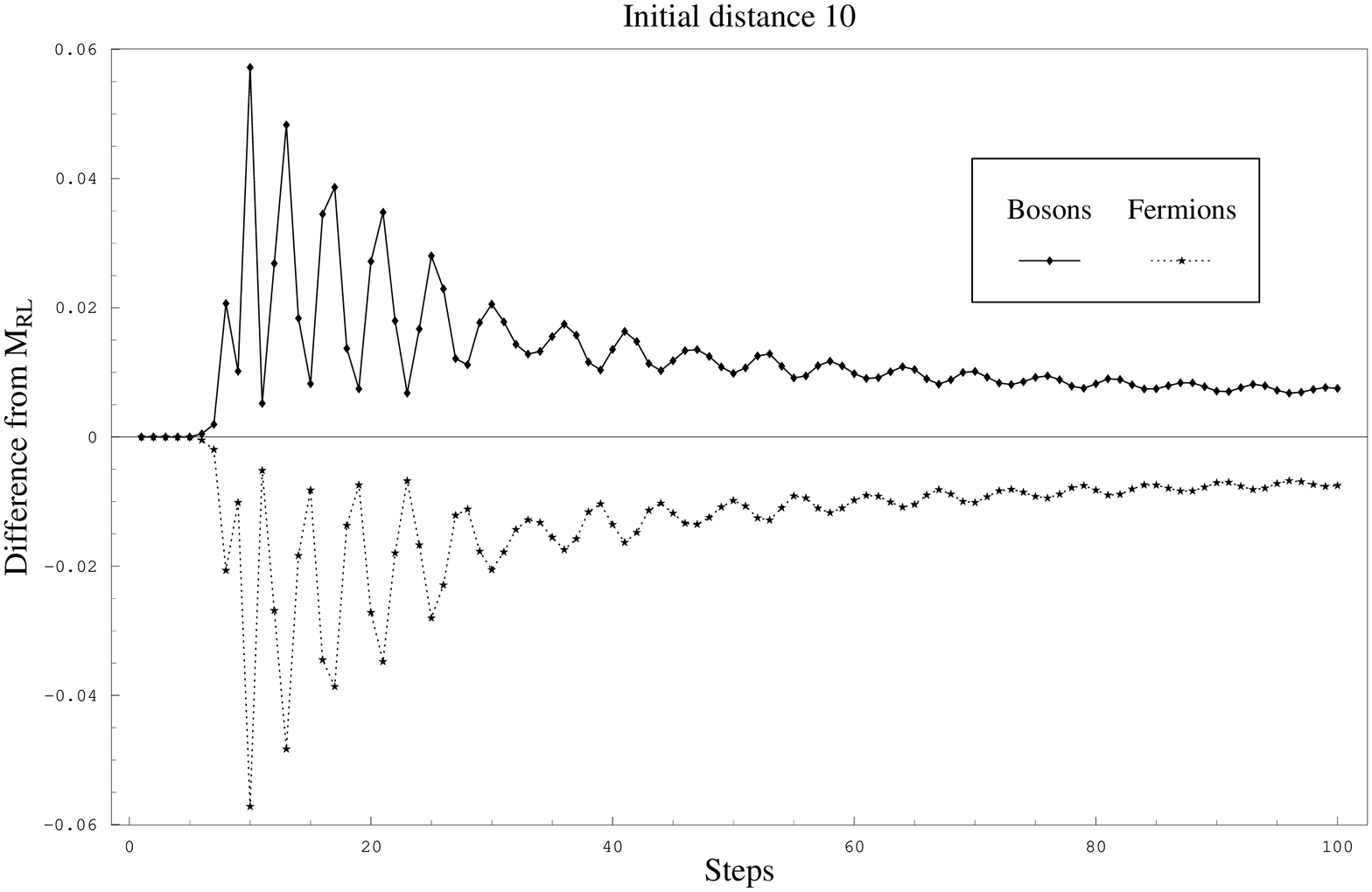}
\end{center}
\caption{Comparison of the meeting probability for bosons,
fermions and distinguishable walkers. The difference in the
meeting probability.}
\end{figure}
From the figure we infer that the peak value is in this case only
slightly changed. Significant differences appear on the long time
scale. The meeting probability is greater for bosons and smaller for
fermions compared to the case of distinguishable walkers. This
behavior can be understood by examining the asymptotic properties of
the expressions (\ref{bfmp}). Numerical evidence indicates that the
meeting probability for bosons has the asymptotic behavior of the
form $\ln(t)/t$. For fermions the decay of the meeting probability
is faster having the form
\begin{equation}
M_F(t,d)\sim\frac{1}{t} .
\end{equation}
The fermion exclusion principle simply works against an enhancement
of the meeting probability.

\section{Conclusions}

We have defined and analyzed the problem of meeting in the quantum
walk on an infinite line with two random walkers. For
distinguishable walkers we have derived analytical formulas for the
meeting probability. The asymptotic behavior following from these
results shows that the meeting probability decays faster but not
quadratically faster than in the classical random walk. This results
in the slower convergency of the overall meeting probability. We
have studied the influence of the entanglement and the
indistinguishability of the walkers on the meeting probability. The
influence is particularly visible for fermions and in the case of
distinguishable walkers for the case of initial entangled singlet
state.

Let us briefly comment on the correspondence between a one
dimensional walk with two random walkers and a two dimensional walk.
As two dimensional walks have been studied by many authors the
possibility of common coin (i.e. a coin which is not a tensor
product) for both walkers has arisen. In the context of one
dimensional walk with two walkers this would mean some kind of
interaction between the two walkers. It would be of interest to find
coins which would attract the walkers and thus lead to an increase
of the meeting probability, or repulsive coins with the opposite
effect. Such interaction would be of infinite range as the walk will
be driven by the same coin independent of the walkers distance. We
can also think about local interactions where the walk is driven by
a common coin for both walkers only when the distance between the
walkers is smaller than some constant or they are at the same
lattice point.

\section{Acknowledgement}

The financial support by GA \v CR 202/04/2101, M\v SMT LC 06001, the
Czech-Hungarian cooperation project (CZ-2/2005,KONTAKT), 
by the Hungarian Scientific Research Fund (T043287 and T049234) and 
EU project QUELE is gratefully acknowledged.

\begin{appendix}

\section{The meeting problem in the classical random walk}

Let us define the meeting problem on the classical level. We
assume two particles which in each step of the process can perform
randomly a step to the left or to the right on a one dimensional
lattice labeled by integers. Initial distance between the two
walkers is $2d$, because for odd initial distance the two walkers
never meet, due to the transitional invariance we can assume that
one walker starts in the vertex labeled by 0 and the other one in
the vertex $2d$. We assume complete randomness, i.e. the
probabilities for the step right or left are equal. We ask for the
probability that the two particles meet again after $t$ steps
either at a certain position $m$ or we might ask for the total
probability to meet (the sum of probabilities at all of the
possible positions). Simple analysis reveals that the probability
to meet at a certain position $m$ equals to
\begin{equation}
M_{cl}(t,m,d) = \frac{1}{2^{2t}} {t\choose \frac{t+m}{2}}
{t\choose \frac{t+m-2d}{2}}. \label{prob1}
\end{equation}
The total probability that the two particles are reunited after
$t$ steps reads
\begin{equation}
M_{cl}(t,d) = \sum\limits^t_{m=2d-t}\frac{1}{2^{2t}} {t\choose
\frac{t+m}{2}} {t\choose \frac{t+m-2d}{2}},\label{prob2}
\end{equation}
which simplifies to
\begin{equation}\label{a1}
M_{cl}(t,d) = \frac{1}{2^{2t}}{2t\choose m+d}.
\end{equation}
This function has a maximum for $t=2d^2$ of value
\begin{equation}
M_{cl}(2d^2, d) = \frac{(4d^2)!}{16^{d^2}(d(2d-1))!(d(2d+1))!}.
\end{equation}
To obtain the asymptotic behavior of the meeting probability we
approximate the one-walker probability distribution by a gaussian
\begin{equation}
P_{cl}(x,t,d) = \frac{1}{\sqrt{\pi t}}\exp(-\frac{(x-2d)^2}{2t}),
\end{equation}
which leads to the following estimate on the meeting probability
\begin{equation}\label{a2}
M_{cl}(t,d) \approx
\int\limits_{-\infty}^{+\infty}P_{cl}(x,t,0)P_{cl}(x,t,d)dx =
\frac{1}{\sqrt{\pi t}} \exp(-\frac{d^2}{t}).
\end{equation}
With the help of this estimation we can simplify the maximal
probability of meeting into
\begin{equation}
M_{cl}(2d^2, d) \approx \frac{1}{\sqrt{2\pi e}d}.
\end{equation}
For a fixed initial distance $d$ we get the long-time
approximation for $t>d^2$
\begin{equation}
M_{cl}(t,d)\approx \frac{1}{\sqrt{\pi t}}(1-\frac{d^2}{t}).
\end{equation}

Finally we study the overall probability that the two walkers will
meet at least once during first $T$ steps, which is given by
\begin{equation}
\overline{M_{cl}}(T,d) = 1-\prod\limits^T_{k=d}{(1-M_{cl}(k,d))}.
\end{equation}
To estimate this function we take the logarithm of
$1-\overline{M_{cl}}(T,d)$ and use the first order of the Taylor
expansion to obtain
\begin{equation}
ln{\left(1-\overline{M_{cl}}(T,d)\right)}=\sum_{k=d}^T
ln{\left(1-M_{cl}(k,d)\right)}\approx -\sum_{k=d}^T M_{cl}(k,d).
\end{equation}
From this with the help of the relation (\ref{a1}) we get the
approximation of the overall meeting probability
\begin{equation}\label{a3}
\overline{M_{cl}}(T,d) \approx
1-\exp\left(-\sum_{k=d}^T\frac{1}{4^k}{2k\choose k+d}\right).
\end{equation}
We estimate the sum in the exponential (\ref{a3}) by an integral
with the help of the formula (\ref{a2})
\begin{eqnarray}\label{a4}
\sum_{k=d}^T\frac{1}{4^k}{2k\choose k+d} &\approx&
\int\limits_0^T\frac{1}{\sqrt{\pi t}} \exp(-\frac{d^2}{t})dt \nonumber \\
 &=&
2\sqrt{\frac{T}{\pi}}\exp(-\frac{d^2}{T})-2d
\textrm{Erfc}(\frac{d}{\sqrt{T}}),
\end{eqnarray}
where $\textrm{Erfc}$ is the complementary error function
\begin{equation}
\textrm{Erfc}(x) = 1-\textrm{Erf}(x) =
1-\frac{2}{\pi}\int\limits_0^x e^{-t^2}dt.
\end{equation}
With the help of the estimation (\ref{a4}) we obtain the
approximation of the overall meeting probability
\begin{equation}
\overline{M_{cl}}(T,d) \approx
1-\exp\left(-2\sqrt{\frac{T}{\pi}}\exp({-\frac{d^2}{T}})\right)\exp\left(2d\textrm{Erfc}(\frac{d}{\sqrt{T}})\right).
\end{equation}

\end{appendix}


\section*{References}
\begin{thebibliography}{99}
\bibitem{crw} R. Burioni and D. Cassi, J. Phys. A 38, R45 (2006)
\bibitem{Aharonov} Y. Aharonov, L. Davidovich, N. Zagury: Phys. Rev. A 48, 1687 (1993)
\bibitem{Kempe} J. Kempe: Contemporary Physics, Vol. 44 (4),
307 (2003)
\bibitem{Shenvi} N. Shenvi, J. Kempe, K. B. Whaley, Phys. Rev. A.
67, 052307 (2003)
\bibitem{Childs} A. M. Childs, R. Cleve, E. Deotto, E. Fahri, S.
Gutman, D. A. Spielman, Proc. 35th Symposium on the Theory of
Computing (ACM Press, New York, 2003), p. 59
\bibitem{qwol} A. Nayak, A. Vishwanath: Quantum walk on a line,
pre-print quant-ph/0010117
\bibitem{knight} P. L. Knight, E. Rold\'an, J. E. Sipe: J. Mod. Opt.
51, 1761 (2004)
\bibitem{carteret1} H. A. Carteret, M. E. H. Ismail, B. Richmond: J.
Phys. A 36, 8775 (2003)
\bibitem{carteret} H. A. Carteret, B. Richmond, N. M. Temme: J. Phys. A 38, 8641
(2005)
\bibitem{NMR} C. A. Ryan, M. Laforest, J. C. Boileau, R. Laflamme: Phys. Rev. A 74, 062317
(2005)
\bibitem{Agarwal} G. S. Agarwal, P.K. Pathak: Phys. Rev. A 72, 033815 (2005)
\bibitem{Hillery} M. Hillery, J. Bergou, E. Feldman: Phys. Rev. A 68, 032314 (2003)
\bibitem{Jeong}
H. Jeong, M. Paternostro, M. S. Kim, Phys. Rev. A 69, 012310 (2004)
\bibitem{Galton} D. Bouwmeester, I. Marzoli, G. P. Karman, W. Schleich and J. P. Woerdman,
Phys. Rev. A 61, 01341 (1999)
\bibitem{Trav} B. C. Travaglione, G. J. Milburn, Phys. Rev. A 65, 032310 (2002)
\bibitem{Dur} W. D\"ur, R. Raussendorf, V. N. Kendon, H. J. Briegel, Phys. Rev. A
66, 052319 (2002)
\bibitem{ion} K. Eckert, J. Mompart, G. Birkl, M. Lewenstein, Phys. Rev. A
72, 012327 (2005)
\bibitem{Man} O. Mandel, M. Greiner, A. Widera, T. Rom, T. W. H\"ansch, I. Bloch,
Phys. Rev. Lett. 91, 010407 (2003)
\bibitem{ctw} E. Farhi, S. Gutman: Phys. Rev. A 58, 915 (1998)
\bibitem{2dqw1} T. D. Mackay, S. D. Bartlett, L. T. Stephenson and
B. C. Sanders: J. Phys. A 35, 2745 (2002)
\bibitem{loc} P. T\"orm\"a, I. Jex, W. P. Schleich Phys. Rev.
A 65, 052110 (2002)
\bibitem{lindberg} K. Lindberg, V. Seshadri, K. E. Shuler, G. H.
Weiss: J. Stat. Phys. 23, 11 (1980)
\bibitem{omar} Y. Omar, N. Paukovic, L. Sheridan, S. Bose: Quantum
walk on a line with two entangled particles, quant-ph/0411065
\bibitem{Tregenna}
B. Tregenna, W. Flanagan, R. Maile, V. Kendon: New Journal of
Physics 5, 83.1 (2003)
\bibitem{abramowitzstegun}
M. Abramowitz, I. A. Stegun: Handbook of Mathematical Functions
with Formulas, Graphs, and Mathematical Tables, Dover Publications
1972
\end{thebibliography}
\end{document}